\documentclass[journal,draftcls,onecolumn]{IEEEtran}
\usepackage[T1]{fontenc}
\usepackage[latin1]{inputenc}
\usepackage{graphicx}
\usepackage{theorem}
\usepackage{amsmath}
\usepackage[caption=false,font=footnotesize]{subfig}
\usepackage{fixltx2e}

\makeatletter

{\theorembodyfont{\rmfamily}
   \newtheorem{The}{{\textbf Theorem}}[section]}
{\theorembodyfont{\rmfamily}
   }
{\theorembodyfont{\rmfamily}
   }
{\theorembodyfont{\rmfamily}
   }
  
\makeatother
\begin{document}

\title{Interleaved Product LDPC Codes}

\author{Marco Baldi,~\IEEEmembership{Member,~IEEE}, Giovanni Cancellieri, and Franco Chiaraluce,~\IEEEmembership{Member,~IEEE}

\thanks{M. Baldi, G. Cancellieri and F. Chiaraluce are with the Dipartimento di Ingegneria dell'Informazione, Università Politecnica delle Marche, Ancona, Italy (e-mail: \{m.baldi, g.cancellieri, f.chiaraluce\}@univpm.it).}}

\maketitle
\begin{abstract}
Product LDPC codes take advantage of LDPC decoding algorithms and
the high minimum distance of product codes.
We propose to add suitable interleavers to improve the waterfall
performance of LDPC decoding.
Interleaving also reduces the number of low weight codewords, that
gives a further advantage in the error floor region.
\end{abstract}

\begin{IEEEkeywords}
Product codes, LDPC codes, Interleavers.
\end{IEEEkeywords}

\section{Introduction}
The current scenario of soft-input soft-output (SISO) decoded error correcting codes is characterized by a huge number of different options, each of them with its own endowment of merits and limitations. Speaking in terms of wide families of codes, classical parallel concatenated turbo codes \cite{Berrou1993} (together with their serial counterpart \cite{Benedetto1998}) generally exhibit easy encoding but rather complex decoding (based on the BCJR algorithm \cite{Bahl1974}). On the contrary, low-density parity-check (LDPC) codes \cite{Gallager} have low decoding complexity, thanks to iterative algorithms working on the Tanner graph, but their encoding complexity can be quadratic in the code length \cite{Richardson2001EfficientEncoding}. Product codes often represent an important tradeoff, as they can exploit a high degree of parallelization both in the encoding and decoding stages. Moreover, they are able to guarantee the value of the minimum distance, that makes them particularly attractive in applications, like optical communications, that require extremely low error rates. Product codes are often designed by using linear block codes as component codes and can be iteratively decoded by using a modified Chase algorithm \cite{Pyndiah1998}, able to provide very good performance especially for high code rate applications. Product codes based on convolutional codes have also been proposed \cite{Gazi2006}. Their component codes exhibit a time invariant trellis structure, so they may be more favorable for implementation than linear block product codes. On the other hand, they require the introduction of interleaving to improve performance and possibly preserve the minimum distance properties.

Much less literature exists, at the authors' knowledge, on the combination of LDPC codes and product codes. Actually, it is well known that long powerful LDPC codes can be constructed by superposition (see \cite{RyanBook} and the references therein). Since the product code can be seen as a special case of superposition, product coding is indeed an effective method to construct irregular LDPC codes \cite{Xu2005}. Till now, however, only a few papers have investigated the features of product LDPC codes. In \cite{Qi2004}, it was shown that they can outperform other LDPC codes constructions in the region of low signal-to-noise ratios. In \cite{Esmaeili2006}, an algorithm was proposed to construct product codes with minimal parity-check matrices, that are expected to improve performance in the waterfall region by increasing the girth. However, such algorithm does not alter the structure of the product code, that, instead, may offer further margins for improving performance. As we show in this letter, such result can be achieved by introducing an interleaver that preserves the multiplicative effect of the product code on the minimum distance.

Regarding the structure of the component codes, in \cite{Xu2005} an Euclidean geometry LDPC code is combined with a single parity-check (SPC) code. In \cite{Thomos2006}, the proposed product code consists of an LDPC code designed through the progressive edge growth (PEG) algorithm \cite{Hu2001PEG}, combined with a Reed-Solomon code, and reveals to be a very efficient solution for error-resilient image transmission. In all cases, a very important issue concerns the need to satisfy the so-called row-column (RC) constraint of the parity-check matrix, that ensures the Tanner graph has girth at least six. If the component codes are both LDPC codes, tighter bounds on the length of local cycles in the product code can also be derived \cite{Esmaeili2006}.
%, with girths $g_a$ and $g_b$, and the parity-check matrix of the product code has a suitable form, then the girth of the LDPC product code is lower bounded by ${\rm min} \{g_a, g_b, 8 \}$ \cite{Esmaeili2006}.

%Once the information bits and the parity bits have been disposed by rows in the encoding matrix (whose definition will be reminded in Section \ref{sec:ProductLDPCCodes}), a common procedure, for linear block codes, is to transmit them by columns.
Classical direct product codes are obtained by placing the information bits in an encoding matrix (that will be better described in Section \ref{sec:ProductLDPCCodes}) and then encoded (first by rows and then by columns, or vice versa). No further interleaving is usually applied. On the contrary, in \cite{Gazi2006}, the effect of different interleavers on the performance of convolutional product codes was investigated, showing that such further randomization can yield significant improvements.

Inspired by the approach in \cite{Gazi2006}, in this letter we extend the application of interleaving to the encoding matrix of product LDPC codes. As component codes we use very simple multiple serially concatenated multiple parity-check (M-SC-MPC) codes, that we have recently introduced \cite{BaldiCL2009}. M-SC-MPC codes have girth at least six and their minimum distance is known or can be easily evaluated through exhaustive enumeration. We apply a suitably designed interleaver that preserves the minimum distance of the product code and satisfies the RC constraint. We show that the new solution provides a significant gain in the waterfall region with respect to the non-interleaved solution, that is the main advantage of the proposed scheme.
%, with performance comparable with that of pseudo-random codes of equal size and rate. 
In addition, interleaving can reduce the multiplicity of low weight codewords, so producing an advantage also in the error floor region.

The letter is organized as follows: Section \ref{sec:ProductLDPCCodes} recalls the definition and properties of product LDPC codes; Section \ref{sec:IntProdLDPCCodes} introduces interleaved product LDPC codes; Section \ref{sec:DesignExamples}
provides some design examples and Section \ref{sec:Conclusion} concludes the letter.

\section{Product LDPC Codes}
\label{sec:ProductLDPCCodes}

\begin{figure}[t]
\begin{centering}
\includegraphics[keepaspectratio]{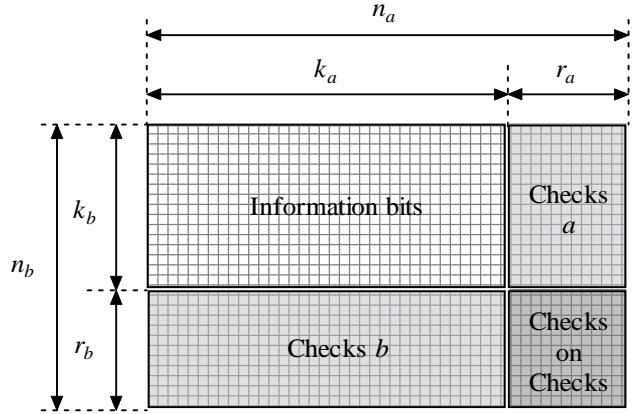}
\caption{\label{fig:PC} Encoding matrix for a direct product code.}
\par\end{centering}
\end{figure}

We focus on the simplest form of product codes, that are bi-dimensional
direct product codes.
In this case, the product code results from two component codes working 
on the two dimensions of a rectangular matrix like that reported in Figure \ref{fig:PC}.
We denote by $\left(n_a, k_a, r_a\right)$ and $\left(n_b, k_b, r_b\right)$ the length, 
dimension and redundancy of the two component codes.
The information bits are written in the top-left $k_b \times k_a$ matrix (marked as ``Information bits''
in the figure) in row-wise order, from top left to bottom right.
When the top-left matrix is filled, the first code, called \textit{row component code}, acts on its rows,
producing a set of $k_b r_a$ checks, that fill the light grey rectangular region 
marked as ``Checks a''.
Then, the second code, called \textit{column component code}, acts on all $n_a$ columns, so producing $k_a r_b$
checks on the information symbols and further $r_a r_b$ checks on checks. The whole matrix has the meaning of an \textit{encoding matrix}.
If the minimum distances of the two component codes are $d_a$ and $d_b$, respectively, 
the product code has minimum distance $d_p=d_a d_b$. The direct product code would be exactly the same if the column component code is applied before the row component code.

Several types of component codes can be used.
SPC codes are often adopted because of their simplicity, but they can yield severe
constraints on the overall code length and rate.
Better results can be obtained with product codes based on Hamming components, that
can achieve very good performance under SISO iterative decoding \cite{Chiaraluce2004}.
%Furthermore, it has been demonstrated that product codes are potentially able to achieve error-free coding with a nonzero code rate (as the number of dimensions increases to infinity) \cite{Elias1954, Rankin2003}.

A parity-check matrix for the direct product code can be obtained as follows.
%from  the parity-check matrices of the two component codes \cite{Baldi2009}.
Let us suppose that the component codes have parity-check matrices $\mathbf{H}_{a}$ and $\mathbf{H}_{b}$, and that $\mathbf{h}_{i,j}$ represents the $j$-th column of $\mathbf{H}_i$, $i = a, b$.
A valid parity-check matrix for the product code having these components is \cite{Baldi2009}:
\begin{equation}
\mathbf{H}_{p}=\left[\begin{array}{c}
\mathbf{H}_{p1}\\
\mathbf{H}_{p2}\end{array}\right],
\label{eq:Hp}
\end{equation}
where $\mathbf{H}_{p1}$ has size $r_a n_b \times n_a n_b$, and $\mathbf{H}_{p2}$ has size $r_b n_a \times n_a n_b$. $\mathbf{H}_{p1}$ is the Kronecker product of an $n_b \times n_b$ identity matrix and $\mathbf{H}_a$, that is,
$\mathbf{I} \otimes \mathbf{H}_a$. This results in a block-diagonal matrix formed by $n_b$ copies of $\mathbf{H}_{a}$,
i.e.,
\begin{equation}
\mathbf{H}_{p1}=\left[\begin{array}{cccc}
\mathbf{H}_{a} 	& \mathbf{0} 			& \cdots & \mathbf{0}\\
\mathbf{0} 			& \mathbf{H}_{a} 	& \cdots & \mathbf{0}\\
\vdots 					& \vdots 					& \ddots & \vdots\\
\mathbf{0} 			& \mathbf{0} 			& \cdots & \mathbf{H}_{a} \end{array}\right],
\label{eq:Hp1}
\end{equation}
where $\mathbf{0}$ represents an $r_a \times n_a$ null matrix.
$\mathbf{H}_{p2}$ is instead a single row of $n_b$ blocks. The $i$-th of these blocks has $n_a$ copies of the $i$-th column of $\mathbf{H}_b$ ($\mathbf{h}_{b,i}$, $i \in \left[1;n_b\right]$) along the main diagonal, while its other symbols are null.

$\mathbf{H}_{p}$ is redundant, since it includes two sets of parity-check constraints representing checks on checks calculated through the two component codes. For this reason, $\mathbf{H}_{p}$ cannot have full rank. 
When the component codes are in systematic form, as in our case, all redundancy bits are positioned at the end of each codeword,
and a full rank parity-check matrix for the product code can be obtained by eliminating the last $r_a r_b$ rows from $\mathbf{H}_{p1}$ or, equivalently, from $\mathbf{H}_{p2}$.
%, in such a way to avoid doubled representation of checks on checks.
In the following, we will choose to eliminate the last $r_a r_b$ rows from $\mathbf{H}_{p1}$, such that it only contains $n_b - r_b = k_b$ copies of $\mathbf{H}_{a}$.
%Such choice ensures full rank for $\mathbf{H}_{p}$, and simplifies the notation.
%that $\mathbf{H}_{p1}$ does not have its last $r_a \cdot r_b$ rows,
%that is, it only contains $n_b - r_b = k_b$ repetitions of $\mathbf{H}_{a}$.
%This gives full rank for $\mathbf{H}_{p}$ and simplifies the notation for interleaved product codes.
An alternative form for $\mathbf{H}_{p2}$ can be obtained if we rearrange its rows by taking, in order, those at the following positions: $1$, $r_b + 1$, $2r_b + 1$, $\ldots$, $\left(n_a - 1\right) r_b + 1$, $2$, $r_b + 2$, $2r_b + 2$, $\ldots$, $\left(n_a - 1\right) r_b + 2$, $\ldots$, $r_b$, $2r_b$, $3r_b$, $\ldots$, $n_a r_b$.
In this case, $\mathbf{H}_{p2}$ can be written as $\mathbf{H}_b \otimes \mathbf{I}$, that is, the Kronecker product of $\mathbf{H}_b$ and an $n_a \times n_a$ identity matrix.
So, according to \cite{Roth2006}, we have:
\begin{equation}
\mathbf{H}_{p}=\left[\begin{array}{c}
\mathbf{I} \otimes \mathbf{H}_a \\
\mathbf{H}_b \otimes \mathbf{I} \end{array}\right].
\label{eq:HpKron}
\end{equation}

If we suppose that the density of symbol $1$ in $\mathbf{H}_a$ and $\mathbf{H}_b$ is $\delta_a$ and $\delta_b$, respectively, it is easy to prove that the density of $\mathbf{H}_{p1}$ is $\delta_a / n_b$, while that of $\mathbf{H}_{p2}$ is $\delta_b / n_a$.
So, even starting from two component codes not having sparse parity-check matrices, the resulting product code can be an LDPC code. 
%Alternative representations of the parity-check matrix can be found, that can achieve even lower density \cite{Esmaeili2006}. For our purposes, however, the density of the parity-check matrix in the form (\ref{eq:Hp}), with $\mathbf{H}_{p1}$ and $\mathbf{H}_{p2}$ as expressed by (\ref{eq:Hp1}) and (\ref{eq:Hp2}), is yet sufficiently low.

Furthermore, it is also possible to verify that the matrix (\ref{eq:Hp}) is free of length-$4$ cycles, provided that the same property holds for the component matrices, $\mathbf{H}_a$ and $\mathbf{H}_b$.
More precisely, it is proved in \cite{Esmaeili2006} that the girth in $\mathbf{H}_{p}$ is lower
bounded by $\min\left\{g_a, g_b, 8\right\}$, where $g_a$ and $g_b$ are the girths in 
$\mathbf{H}_a$ and $\mathbf{H}_b$, respectively.
So, the codes obtained as bi-dimensional product codes can be effectively decoded by means of LDPC decoding algorithms, like the well-known Sum-Product Algorithm (SPA) \cite{Hagenauer1996}, acting on the code Tanner graph. Compared to classical turbo product code decoding techniques, that exploit iterative decoding of the component codes, SPA achieves the same or better performance, but with lower complexity \cite{Baldi2009}.
%We will give some examples in this sense in the next section.

\section{Interleaved Product LDPC Codes}
\label{sec:IntProdLDPCCodes}

A common solution to improve the convergence of iterative soft-decision decoding algorithms
is to insert an interleaver between two (or more) concatenated component codes.
Interleaving is crucial in the design of turbo codes, and it has also been exploited in the 
design of turbo product codes based on convolutional codes \cite{Gazi2006}.

We are interested in the use of column-interleavers, that are able to preserve 
the minimum distance of the product code by interleaving only one of the two component 
codes.
In other terms, a column-interleaver only permutes the elements within each row of the encoding matrix.
%reported in Fig. \ref{fig:PC}.
Since the interleaver acts after row encoding, the effect of the row component code is unaltered and, before column encoding, at least $d_a$ columns contain a symbol $1$.
It follows that the code minimum distance remains $d_p = d_a d_b$ \cite{Gazi2006}.

A valid parity-check matrix for the column-interleaved product code can be obtained starting
from (\ref{eq:HpKron}) and considering that:
\begin{equation}
\begin{array}{rcl}
\mathbf{H}_{p2} & = & \mathbf{H}_b \otimes \mathbf{I} \\
& = & \left[\mathbf{h}_{b,1} | \mathbf{h}_{b,2} | \ldots | \mathbf{h}_{b,n_b} \right] \otimes \mathbf{I} \\
& = & \left[\mathbf{h}_{b,1} \otimes \mathbf{I} | \mathbf{h}_{b,2} \otimes \mathbf{I} | \ldots | \mathbf{h}_{b,n_b} \otimes \mathbf{I} \right]. \\
\end{array}
\label{eq:HpKronColumns}
\end{equation}
%where $\mathbf{h}_{b,i}$ denotes the $i$-th column of matrix $\mathbf{H}_b$, $i \in \left[1;n_b\right]$.
Let us introduce a vectorial Kronecker product operator ($\overline{\otimes}$) that works \textit{column-wise}. 
%In order to compute its result, we repeat the matrix on the right of the operator as many times as the number of columns in the matrix on the left.
Given a $x \times y$ matrix $\mathbf{A}$ and a $v \times wy$ matrix $\mathbf{B}$, $\mathbf{C} = \mathbf{A} \overline{\otimes} \mathbf{B}$ is a $xv \times wy$ matrix. The $i$-th group of $w$ columns of $\mathbf{C}$,
$i = 1,\ldots,y$, is obtained by starting from the $i$-th column of $\mathbf{A}$ and multiplying each element by the matrix
formed by the $i$-th group of $w$ columns of $\mathbf{B}$.
By using this operator, \eqref{eq:HpKronColumns} can be rewritten as follows:
\begin{equation}
\mathbf{H}_{p2} = \mathbf{H}_b \overline{\otimes} \ \overline{\mathbf{I}} = \mathbf{H}_b \overline{\otimes} \left[\mathbf{I} | \mathbf{I} | \ldots | \mathbf{I} \right], \\
\label{eq:HpVectKron}
\end{equation}
where the right operand matrix is a row of $n_b$ identity matrices, each with size $n_a$.
By the explicit computation of \eqref{eq:HpVectKron}, it can be shown that $\mathbf{H}_{p2}$ contains $n_a$ copies of $\mathbf{H}_b$, each copy having its elements spread within $\mathbf{H}_{p2}$.
In fact, each element of $\mathbf{H}_b$ is replaced by an $n_a \times n_a$ identity matrix, according to \eqref{eq:HpVectKron},
so all the elements of $\mathbf{H}_b$ are repeated $n_a$ times within $\mathbf{H}_{p2}$.
More in detail, a first copy of $\mathbf{H}_b$ involves the codeword bits at 
positions $1$, $n_a + 1$, $2n_a + 1$, \ldots, $(n_b-1)n_a + 1$; a second copy of $\mathbf{H}_b$
involves the codeword bits at position $2$, $n_a + 2$, $2n_a + 2$, \ldots, $(n_b-1)n_a + 2$ and so on.

Let us consider the array $\overline{\mathbf{P}} = \left[\mathbf{P}_1 | \mathbf{P}_2 | \ldots | \mathbf{P}_{n_b} \right]$ of $n_b$ permutation matrices, each with size $n_a$. The permutation matrix $\mathbf{P}_j$, $j \in \left[1;n_b\right]$, can be described through the set $\Pi^j = \left\{\pi_1^j, \pi_2^j, \ldots, \pi_{n_a}^j \right\}$, in which $\pi_i^j$ is the column index of the symbol $1$ at row $i$.
%in matrix $\mathbf{P}_j$.
It follows from the definition of permutation matrix that $\Pi^j$ has no duplicate elements.

\begin{The}
\label{th:HpP}
Given a product code with parity-check matrix $\mathbf{H}_p$ in the form \eqref{eq:HpKron}, the application of a column-interleaver transforms $\mathbf{H}_p$ into:
\begin{equation}
\mathbf{H}_{p}^P=
\left[\begin{array}{c}
\mathbf{I} \otimes \mathbf{H}_a \\
\mathbf{H}_b \overline{\otimes} \ \overline{\mathbf{P}}
\end{array}\right],
\label{eq:HpP}
\end{equation}
where $\overline{\mathbf{P}}$ is the array of permutations applied to the $n_b$
rows of the encoding matrix.
%(see Fig. \ref{fig:PC}).
\end{The}
\begin{IEEEproof}
As described above, a column-interleaver only permutes the column component code. Since $\mathbf{H}_{p1} = \mathbf{I} \otimes \mathbf{H}_a$ describes the row component code, it remains unchanged after interleaving.
%Explicitly, this means that it consists of $n_b - r_b = k_b$ repetitions of $\mathbf{H}_{a}$.
%On the other hand, in $\mathbf{H}_{p2} = \mathbf{H}_b \otimes \mathbf{I}$, the $n_a$  repetitions of $\mathbf{H}_b$ are transformed such that the first of them involves the codeword bits at  positions $1$, $n_a + 1$, $2n_a + 1$, \ldots, $(n_b-1)n_a + 1$; the second repetition of $\mathbf{H}_b$  involves the bits at position $2$, $n_a + 2$, $2n_a + 2$, \ldots, $(n_b-1)n_a + 2$ and so on.
By replacing $\mathbf{H}_{p2}$ with $\mathbf{H}_{p2}^P = \mathbf{H}_b \overline{\otimes} \ \overline{\mathbf{P}}$, the first copy
of $\mathbf{H}_b$ within $\mathbf{H}_{p2}^P$ involves the codeword bits at positions $\pi_1^1$, $n_a + \pi_1^2$, $2n_a + \pi_1^3$, 
\ldots, $(n_b - 1)n_a + \pi_1^{n_b}$; the second copy of $\mathbf{H}_b$ within $\mathbf{H}_{p2}^P$ involves the codeword 
bits at positions $\pi_2^1$, $n_a + \pi_2^2$, $2n_a + \pi_2^3$, \ldots, $(n_b - 1)n_a + \pi_2^{n_b}$
and so on. The indexes of the codeword bits involved in each copy of $\mathbf{H}_b$ within $\mathbf{H}_{p2}^P$ are all distinct, since 
$\pi_i^j$, $\forall i,j$, takes values in the range $\left[1;n_a\right]$.
Furthermore, all codeword bits involved in the same copy of $\mathbf{H}_b$ come from different rows
of the encoding matrix. More precisely, the $m$-th codeword bit involved in the $q$-th copy
of $\mathbf{H}_b$ is at position $(m-1)n_a + \pi_q^m$.
Since this value is between $(m-1)n_a + 1$ and $m n_a$, the bit comes from the $m$-th row of the encoding matrix.
Finally, for the properties of permutation matrices, each $\Pi^j = \left\{\pi_1^j, \pi_2^j, \ldots, \pi_{n_a}^j \right\}$,
$j \in \left[1;n_b\right]$, does not contain duplicate elements; so, each codeword bit is only involved in one copy of $\mathbf{H}_b$.
This proves that $\mathbf{H}_{p}^P$ describes the product code after application of the column-interleaver.
\end{IEEEproof}

\begin{figure}[t]
\begin{centering}
\includegraphics[keepaspectratio]{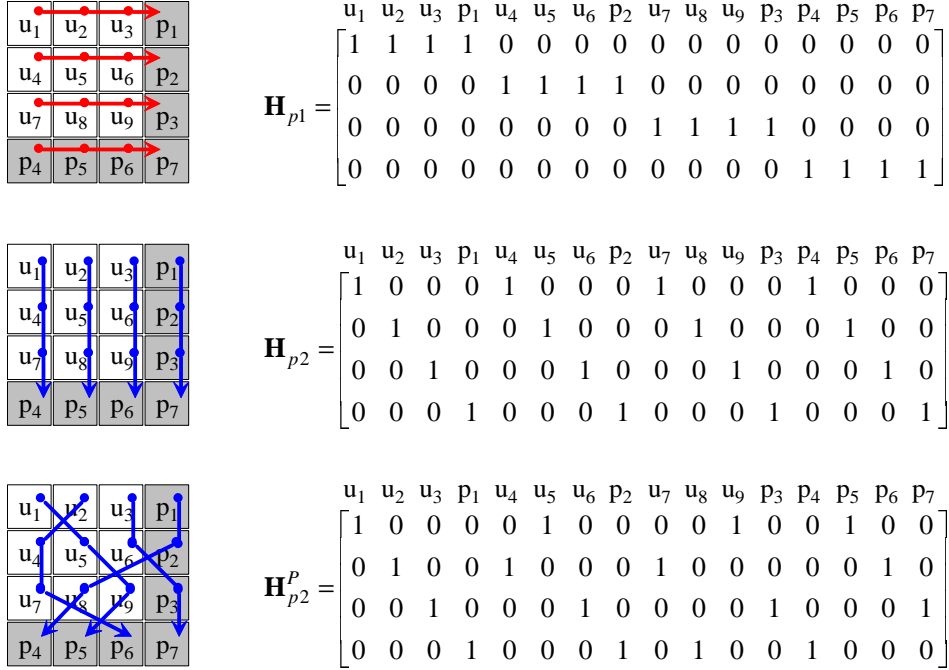}
\caption{\label{fig:didactic} Example of product code and its column-interleaved version. The component codes are both SPC(4, 3). The applied permutations are described by $\Pi^1 = \left\{1, 2, 3, 4 \right\}, \Pi^2 = \left\{2, 1, 3, 4 \right\}, \Pi^3 = \left\{3, 1, 4, 2 \right\}, \Pi^4 = \left\{2, 3, 4, 1 \right\}$. $\mathbf{H}_{p1}$ is shown before eliminating the last row (required to have a full rank parity-check matrix).} 
\par\end{centering}
\end{figure}

A tutorial example of product code and its column-interleaved version is shown in Fig. \ref{fig:didactic}.

Theorem \ref{th:HpP} establishes a method for the design of the parity-check matrix of a product
code in which a column-interleaver is applied.
We will denote such product codes as \textit{interleaved product codes} in the following.
Since we are interested in product codes that are also LDPC codes, to be decoded through LDPC 
decoding algorithms, it is important that the corresponding Tanner graph is free of
short cycles.
To this purpose, we can extend the results obtained in \cite{Esmaeili2006} as follows.

\begin{The}
\label{th:Cycles}
The parity-check matrix of an interleaved product code, $\mathbf{H}_{p}^P$, in the form
\eqref{eq:HpP}, has local cycles with length $\geq \min\left\{g_a, g_b, 8\right\}$, where 
$g_a$ and $g_b$ are the girths in $\mathbf{H}_a$ and $\mathbf{H}_b$, 
respectively.
\end{The}
\begin{IEEEproof}
As in \cite{Esmaeili2006}, we define the number of connections between two matrices with
equal size as the number of columns in which both matrices have at least a symbol $1$.
Within the parity-check matrix of a product code, it can be observed that each copy of 
$\mathbf{H}_a$ has only one connection with any copy of $\mathbf{H}_b$ \cite{Esmaeili2006}.
More precisely, we observe that, within $\mathbf{H}_{p}$ in the form \eqref{eq:Hp},
the $i$-th column of each copy of $\mathbf{H}_a$ is connected only with the $i$-th
copy of $\mathbf{H}_b$.
Due to column-interleaving, within $\mathbf{H}_{p}^P$, having the form \eqref{eq:HpP},
the $i$-th column of the $j$-th copy of $\mathbf{H}_a$ is connected with the
$\pi_i^j$-th copy of $\mathbf{H}_b$.
Since $\Pi^j = \left\{\pi_1^j, \pi_2^j, \ldots, \pi_{n_a}^j \right\}$ does not contain duplicate elements, the $j$-th copy of $\mathbf{H}_a$, $\forall j \in [1; n_b]$, has only one connection with each copy of 
$\mathbf{H}_b$.
%The same holds for each repetition of $\mathbf{H}_a$.
So, the same arguments
used in \cite{Esmaeili2006} for a direct product code apply, 
and this proves the theorem.
\end{IEEEproof}

Based on Theorem \ref{th:Cycles}, the proposed class of interleaved product codes can be seen as LDPC codes
with Tanner graphs suitable for the application of decoding algorithms based on belief propagation.

An important task is to design the array of permutation matrices $\overline{\mathbf{P}}$ in such a way as to have a Tanner graph with good properties for decoding.
%We must observe that the choice of the array of permutations $\overline{\mathbf{P}}$ is not unique, and a valuable task is to design it in such a way as to optimize the code Tanner graph.
%The application of column-interleaving gives a further degree of freedom in the design of
%the parity-check matrix, that is, the choice of the set of permutations $\overline{\mathbf{P}}$,
%while allowing to preserve the code minimum distance.
%So, in order to improve the performance of the LDPC decoder, the choice of $\overline{\mathbf{P}}$
%can be made through algorithms aimed at optimizing the code Tanner graph, as the PEG algorithm \cite{Hu2001PEG}.
To this goal, we have developed two modified versions of the PEG algorithm \cite{Hu2001PEG}.
%Both of them are aimed at assigning the array of permutations $\overline{\mathbf{P}}$
%of an interleaved product code by increasing, as much as possible, the length of local cycles in its
%associated Tanner graph.
Both of them aim at selecting, for an interleaved product code, an array of permutation matrices $\overline{\mathbf{P}}$
which maximizes the length of local cycles.
% in its associated Tanner graph.
The array of permutation matrices $\overline{\mathbf{P}}$, so designed, is then used to obtain $\mathbf{H}_{p2}^P = \mathbf{H}_b \overline{\otimes} \ \overline{\mathbf{P}}$.
%More precisely, the modified PEG algorithms realize a matrix $\mathbf{H}_{p}^P$,
%in the form \eqref{eq:HpP}, at first by filling $\mathbf{H}_{p1} = \mathbf{I} \otimes \mathbf{H}_a$ with $k_b$ repetitions of $\mathbf{H}_a$,
%then by assigning the permutations in $\mathbf{H}_{p2}^P = \mathbf{H}_b \overline{\otimes} \ \overline{\mathbf{P}}$
%according with the PEG principle. For this purpose, the original PEG algorithm has 
%been modified in such a way to:
The original PEG algorithm has been modified in order to:
\begin{itemize}
\item insert edges only in those $n_a \times n_a$ blocks of $\mathbf{H}_{p2}^P$ that correspond to a symbol $1$ in $\mathbf{H}_b$;
\item verify the permutation matrix constraint by inserting only a symbol $1$ in any row and column of each $n_a \times n_a$ block;
\item apply the vectorial Kronecker product, so that the same permutation matrix appears in all blocks along each column of $\mathbf{H}_{p2}^P$.
\end{itemize}
This allows to obtain a matrix $\mathbf{H}_{p2}^P$ that, together with $\mathbf{H}_{p1} = \mathbf{I} \otimes \mathbf{H}_a$ as in \eqref{eq:HpP}, forms a valid parity-check matrix for the interleaved product code.

The two versions of the modified PEG algorithm differ in the type of permutation matrices they use. The first version only uses circulant permutation matrices. This further constraint reduces the margin for local cycles optimization but produces a structured $\mathbf{H}_{p}^P$. As well known, a structured matrix is advantageous in regard to the hardware implementation of encoders and decoders. The second version instead uses general permutation matrices.
%Based on these constraints, both versions of the modified PEG algorithm produce valid parity-check
%matrices for the interleaved product code.
%However, the first modified version includes another constraint, that is, it only inserts
%circulant permutation matrices.
%Such further constraint obviously reduces the margin for local cycles optimization, but preserves
%the structured nature of $\mathbf{H}_{p}^P$.
%In fact, in this case, identity matrices in $\mathbf{H}_{p}$ are replaced with circulant permutation
%matrices, that still have a number of advantages concerning the hardware implementation of encoders
%and decoders.
%The second version of the modified algorithm does not include such constraint,
%and therefore produces generic permutation matrices.
This choice increases the randomization level in $\mathbf{H}_{p}^P$ and provides further margins
for optimization, but its implementation in hardware may be more complex than for structured matrices.
%despite the loss of its inner structure.

%In the next section we show some design examples of product codes and interleaved product codes, obtained through both the described versions of the modified PEG algorithm.

\section{Design Examples}
\label{sec:DesignExamples}

We provide some design examples of product LDPC codes and their column-interleaved versions by focusing on two values of code rate, namely, $R = 2/3$ and $R = 3/4$.
%For the former value of code rate, we have used, for both components of the product code, an
%LDPC code designed through the PEG algorithm \cite{Hu2001PEG}.
%It has length $n_a = n_b = 56$ and dimension $k_a = k_b = 40$. Its parity-check matrix is
%in lower triangular form and encoding is systematic, with the redundancy part representing
%the last $r_a = r_b = 16$ bits in each codeword.
%The minimum distance of such code, estimated through an exhaustive search, is $d = 4$
%and its corresponding multiplicity is $M_4 = 35$.
For the case of $R = 2/3$, we have used, for both components of the product code, an
M-SC-MPC code with $M=2$ and $r_j = \left[9,10\right]$ \cite{BaldiCL2009}.
It has length $n_a = n_b = 100$ and dimension $k_a = k_b = 81$. The parity-check 
matrix of each component code is in lower triangular form and encoding is systematic, with the 
$r_a = r_b = 19$ redundancy bits in the rightmost part of each codeword.
Through an exhaustive search, the minimum distance $d = 4$ has been found.
Its corresponding multiplicity is $M_4 = 2025$.

Similarly, for $R = 3/4$, we have used as components two identical
M-SC-MPC codes with $M=2$, $r_j = \left[13,14\right]$, $n_a = n_b = 196$ and $k_a = k_b = 169$.
An exhaustive search has reported that the minimum distance is $d = 4$ and its multiplicity
is $M_4 = 8281$.

For both values of the code rate, we have designed a product code with parity-check matrix in the
form \eqref{eq:Hp}. These two product codes are denoted as PC in the following. They have 
$\left(n,k\right) = \left(10000, 6561\right)$ and $\left(n,k\right) = \left(38416, 28561\right)$, respectively.
Moreover, we have designed two column-interleaved product codes for each value of code rate, by
applying the two modified versions of the PEG algorithm described in the previous section.
Obviously, they have exactly the same length and rate as their corresponding product codes, but their
parity-check matrices are different.
The first interleaved product code, denoted as iPC-CP,
has been designed trough the modified PEG algorithm with the constraint of using
only circulant permutation matrices.
The second interleaved product code, denoted as iPC-RP, has been obtained
through the modified PEG algorithm that uses generic permutation matrices.
%, based on a constrained random criterion.

The performance of the considered codes has been assessed by simulating 
Binary Phase Shift Keying transmission over the Additive White Gaussian Noise channel.
LDPC decoding has been performed through the SPA with Log-Likelihood Ratios.
For each value of the energy per bit to noise power spectral density ratio ($E_b/N_0$),
a value of Bit Error Rate (BER) and Frame Error Rate (FER) has been estimated through
a Montecarlo simulation, waiting for the occurrence of a sufficiently high number of erred frames, in order
to reach a satisfactory confidence level. The union bound for the product code, noted as PC UB, has been 
used as a reference.

%Fig. \ref{fig:R12} reports the simulation results for codes with rate $1/2$. 
%As we observe from the figure, the introduction of a column-interleaver has
%the effect of improving the product code performance.
%In terms of coding gain, the two interleaved product codes have an advantage
%of about $0.5$ dB with respect to the product code.
%There is no relevant difference between the performance of the iPC-CP code
%and that of the iPC-RP one.

%\begin{figure}
%\begin{centering}
%\subfloat[]{\includegraphics[width=43mm]{R12_BER}}
%\subfloat[]{\includegraphics[width=43mm]{R12_FER}}
%\par\end{centering}
%\caption{(a) BER and (b) FER for $(3136, 1600)$ product and interleaved product codes.
%\label{fig:R12}}
%\end{figure}

\begin{figure}
\begin{centering}
\subfloat[]{\includegraphics[width=70mm]{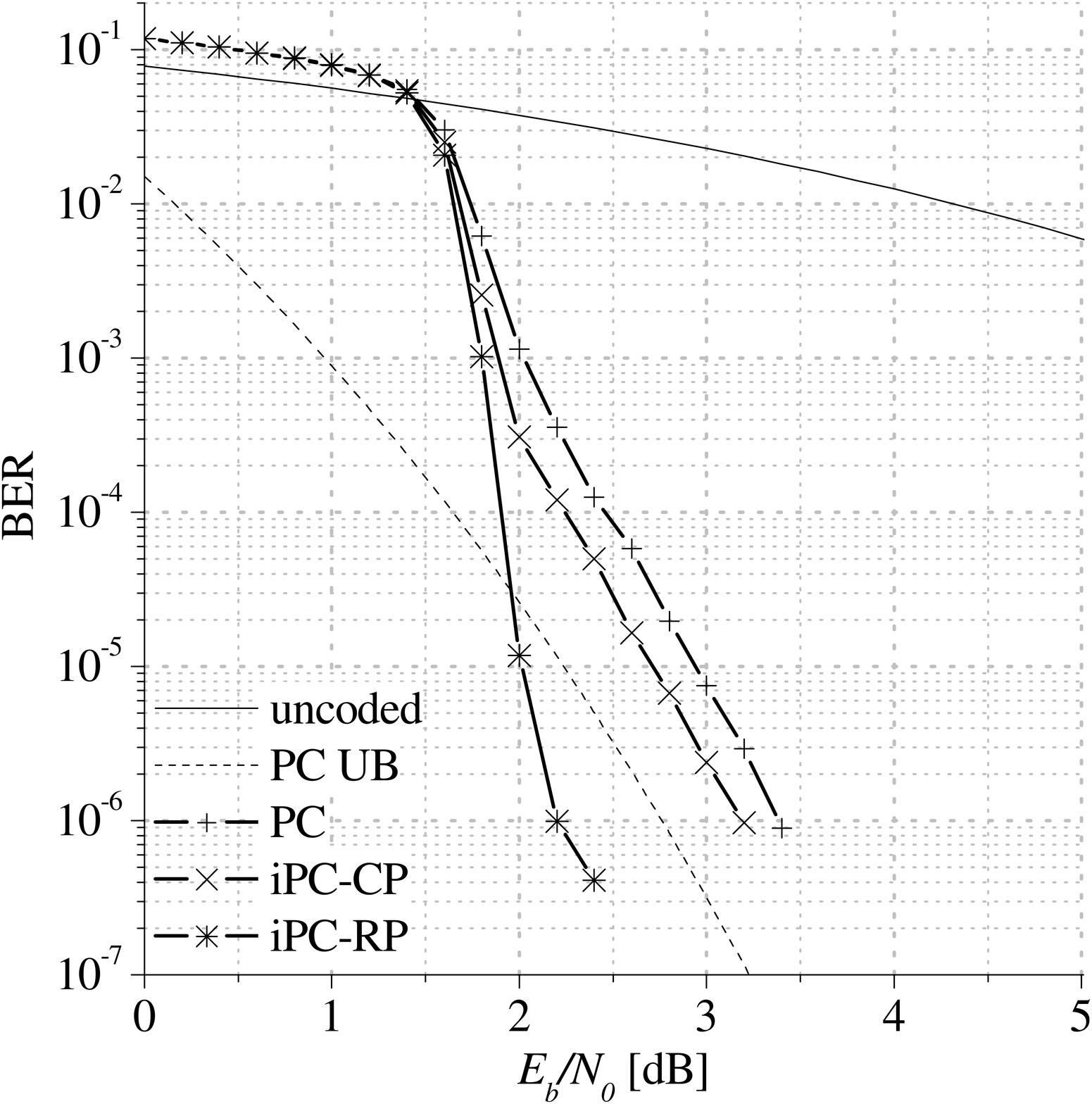}}
\subfloat[]{\includegraphics[width=70mm]{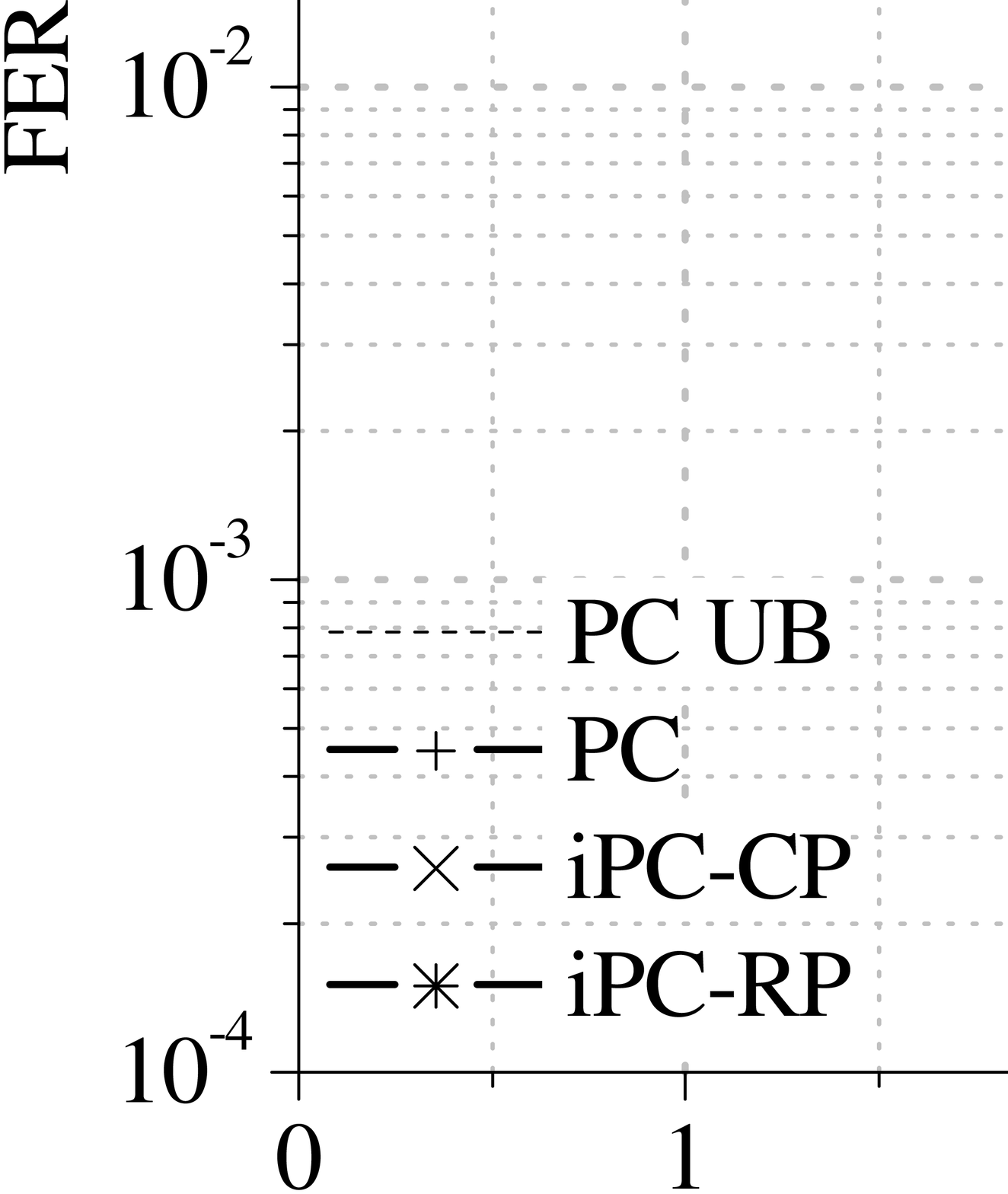}}
\par\end{centering}
\caption{(a) BER and (b) FER for $(10000, 6561)$ product and interleaved product codes.
\label{fig:R23}}
\end{figure}

Fig. \ref{fig:R23} shows the simulation results for codes with rate $2/3$.
We observe that, in this case, the iPC-CP code achieves an improvement in coding 
gain of about $0.2$ dB with respect to the product code.
The iPC-RP code outperforms the iPC-CP code, and it is able to reach a significant 
improvement, by more than $1$ dB, with respect to the product code. Another interesting remark is that the curves for the iPC-RP code intersect the PC UB curves.

We conjecture that such improvement is due to a reduction in the multiplicity of low weight codewords.
The product code, in fact, has a rather high number of minimum weight codewords, that is, $M_4 = 2025^2$ for the present case.
The effect of column-interleaving is to reduce such multiplicity, particularly for the iPC-RP code, that has been designed with no constraint on the permutation matrices.
%as it suffers less constraints than the iPC-CP code. This is also demonstrated by the fact that the curves for the iPC-RP code intersect the PC UB curves.

To verify this conjecture, we have considered the simple case of an $\left(n,k\right) = \left(144, 25\right)$ product code obtained by using, as component codes, two identical M-SC-MPC codes with $M=2$ and $r_j = \left[3,4\right]$, having length $n_a = n_b = 12$ and dimension $k_a = k_b = 5$. Being very small, these codes permit us to analyze the whole weight spectrum of the product code and its interleaved versions. The first terms of the weight spectra are reported in Table \ref{tab:spectrum}.
Though referred to small codes, the results show that interleaving reduces the multiplicity of minimum weight codewords (it passes from 64 to 40 for both the interleaved codes).
We also notice, for the interleaved codes, the appearance of weights that were absent in the product code.
Despite this, we observe that the multiplicities of the lowest weights for the interleaved codes are generally lower than those
for the direct product code.
This effect, that is similar to the spectral thinning occurring in turbo codes \cite{Perez1996}, is most evident for the iPC-RP code.

\begin{table*}[ht]
\caption{First terms of the weight spectrum for a (144, 25) product LDPC code and its interleaved versions}

\begin{centering}
\begin{tabular}{cccc}
\hline
weight & PC & iPC-CP & iPC-RP \\
\hline
$16$ & $64$ & $40$ & $40$ \\
$20$ & - & $8$ & $6$ \\
$22$ & - & $2$ & - \\
$24$ & $246$ & $143$ & $116$ \\
$26$ & - & $23$ & $24$ \\
$28$ & $504$ & $317$ & $330$ \\
$30$ & $392$ & $244$ & $211$ \\
$32$ & $1262$ & $831$ & $719$ \\
$...$ & $...$ & $...$ & $...$ \\
\hline
\end{tabular}
\par\end{centering}
\label{tab:spectrum} 
\end{table*}

\begin{figure}
\begin{centering}
\subfloat[]{\includegraphics[width=70mm]{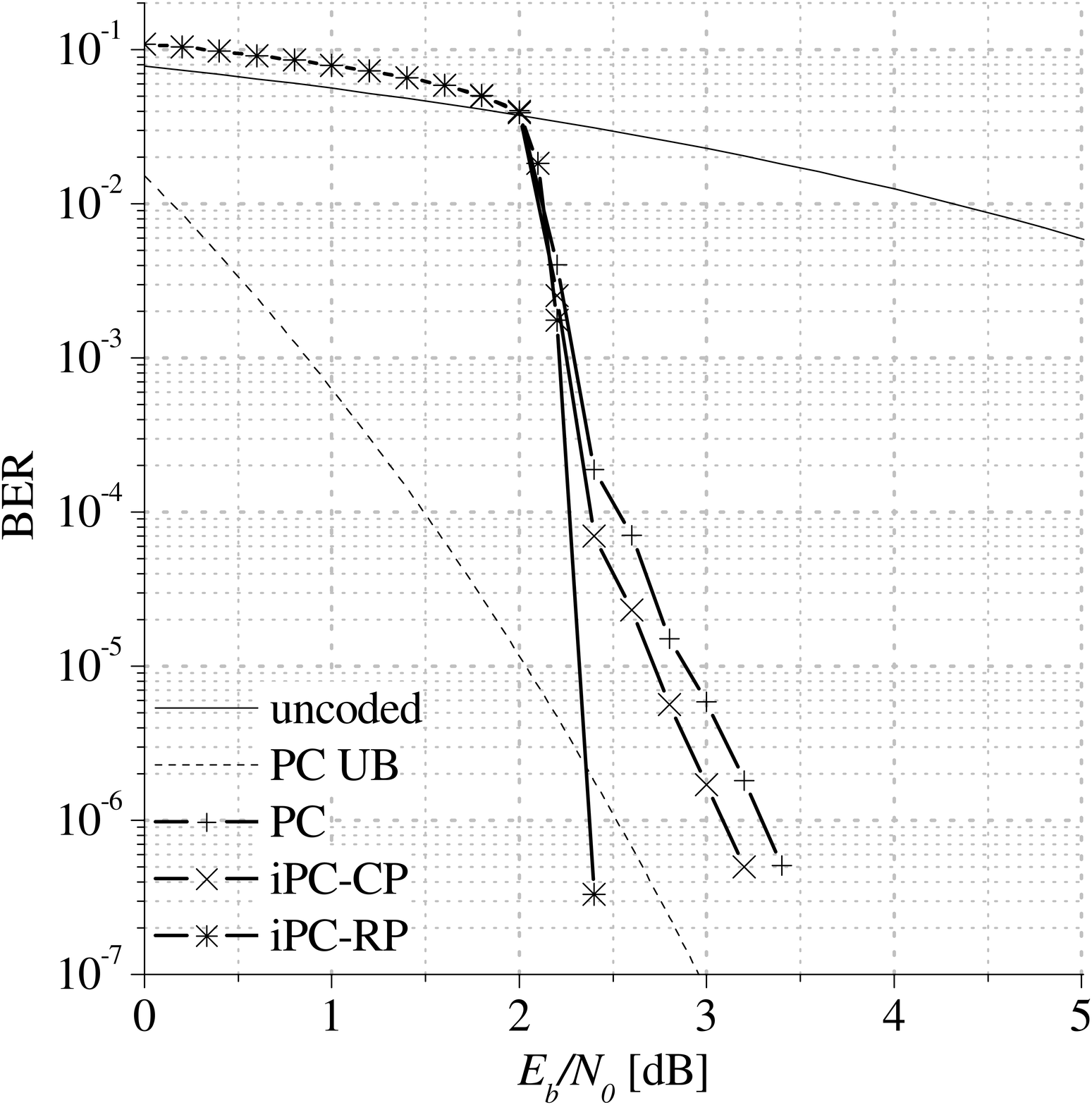}}
\subfloat[]{\includegraphics[width=70mm]{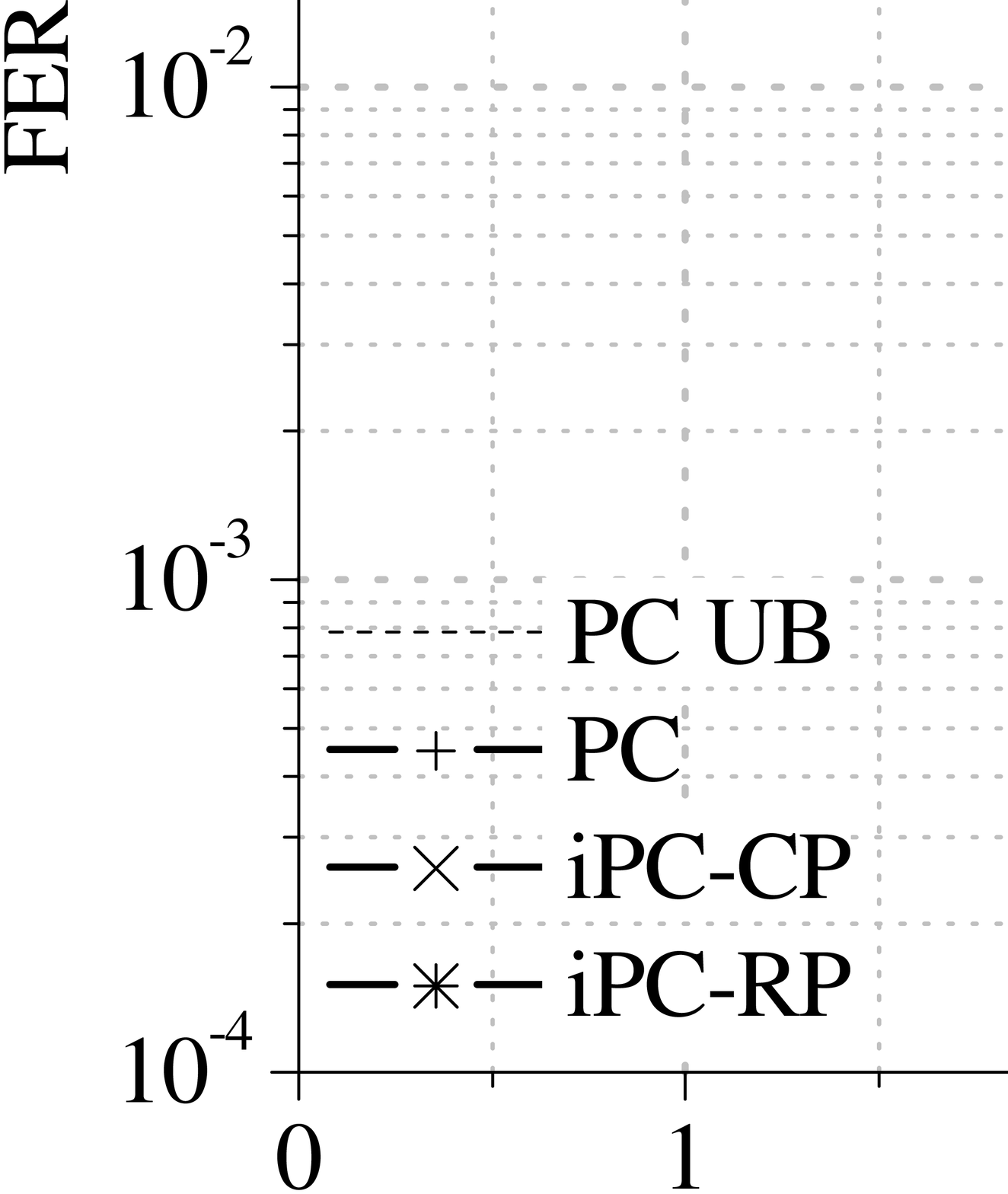}}
\par\end{centering}
\caption{(a) BER and (b) FER for $(38416, 28561)$ product and interleaved product codes.
\label{fig:R34}}
\end{figure}

The advantage of interleaving is even more evident for longer codes, with rate $3/4$, whose simulated
performance is shown in Fig. \ref{fig:R34}.
In this case, the multiplicity of low weight codewords in the direct product code is even higher, and the advantage
due to the PEG-based random interleaver more remarkable.
Interleaving based on circulant matrices gives an improvement of about $0.2$ dB with
respect to the classical product code. By using an interleaver based on generic permutation 
matrices, the gain exceeds $1$ dB and, differently from the code with rate $2/3$,
no error floor effect is observed in the explored region of $E_b/N_0$ values. Obviously, as interleaving does not increase the value of the minimum distance, for high signal-to-noise ratio the error rate curves for direct and interleaved product codes
must assume the same slope. However, the spectral thinning effect of interleaving determines the slope change at significantly smaller error rate values.

\begin{figure}
\begin{centering}
\includegraphics[width=70mm]{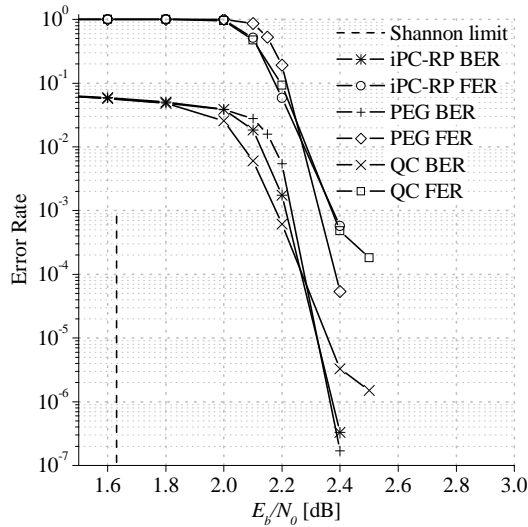}
\par\end{centering}
\caption{Performance comparison, for $R = 3/4$, between the interleaved product codes and conventional, QC and PEG, LDPC codes.}
\label{fig:C34}
\end{figure}

Finally, in order to further assess the performance of the proposed class of codes, we have compared the obtained error rate curves with those of structured and unstructured LDPC codes, not in the form of product codes. An example is shown in Fig. \ref{fig:C34} for $R = 3/4$. The structured code is Quasi-Cyclic (QC), and has been obtained by extending to length $n = 38400$ the design approach used for the rate $3/4$ ``B'' code of the IEEE 802.16e standard \cite{802.16e}. The unstructured code, instead, has exactly the same parameters of the rate $3/4$ product code, and has been designed through the PEG algorithm. Also the Shannon limit has been plotted as a reference. Though all codes exhibit a gap from the best result theoretically achievable, their performance is very similar.
The BER curve of the interleaved product LDPC code is between those of the PEG code (that achieves the best performance)
and the QC code (that shows good waterfall behavior, but also the appearance of an error floor effect).
Thus, it is confirmed that the interleaved product LDPC codes, based on very simple M-SC-MPC code components, do not suffer a performance loss with respect to other state-of-the-art LDPC design solutions.
This is even more evident if we consider that some margin for further improving the performance of the proposed class of interleaved product codes may exist, because of the degrees of freedom in the constrained random behavior of the PEG algorithm.

\section{Conclusion \label{sec:Conclusion}}

%We have shown how a column-interleaver can be introduced in a product LDPC code
%and that the resulting parity-check matrix has a very simple structure, derived
%from that of the classic product code.
%We have proved that interleaved product LDPC codes, so obtained, are still LDPC
%codes, and their associated Tanner graph is free of short cycles.

%We have proposed a design approach for product LDPC codes interleavers, based on
%the PEG algorithm, that aims at optimizing the length of local cycles in the
%Tanner graph while preserving the multiplicative effect of the product code on
%the minimum distance.

%The proposed design approach is able to produce either structured parity-check matrices,
%formed by circulant permutation matrices, or non-structured ones, based on random
%permutation matrices.
%Both them are able to improve the performance of classic product LDPC codes.
%For high code rate and minimum weight multiplicity, the non-structured approach
%is able to achieve a significant improvement in coding gain and to outperform the
%structured one.

We have shown that interleaved product LDPC codes can have very good performance both in the error floor region, where they benefit by a large (and guaranteed) minimum distance value, and in the waterfall region, through the design of suitable column-interleavers.
We have proposed two different versions of a modified PEG algorithm for the design of column-interleavers: the first one uses circulant permutation matrices while the second one exploits generic permutation matrices. The first version preserves the structured nature of the parity-check matrix that, instead, is lost with the second version. As a counterpart, the use of generic permutation matrices gives the best performance, mostly because of the spectral thinning effect.

We wish to stress that the column-interleaver design is not critical, in the sense that, following the proposed procedure, many different permutations can be found with similar performance.
%Even more, we can say that the interleavers used in the considered examples are not optimized, so that further margins for improvement may exist.
Although interleaving can be applied to product LDPC codes of any length and rate, our simulations show that the coding gain advantage is more pronounced for long codes with rather high rates. Once again, this can be explained in terms of the spectral thinning effect, which provides the rationale of the performance improvement we have found.

\bibliographystyle{IEEEtran}
\bibliography{Baldi_CL2010}

\begin{thebibliography}{10}
\providecommand{\url}[1]{#1}
\csname url@samestyle\endcsname
\providecommand{\newblock}{\relax}
\providecommand{\bibinfo}[2]{#2}
\providecommand{\BIBentrySTDinterwordspacing}{\spaceskip=0pt\relax}
\providecommand{\BIBentryALTinterwordstretchfactor}{4}
\providecommand{\BIBentryALTinterwordspacing}{\spaceskip=\fontdimen2\font plus
\BIBentryALTinterwordstretchfactor\fontdimen3\font minus
  \fontdimen4\font\relax}
\providecommand{\BIBforeignlanguage}[2]{{%
\expandafter\ifx\csname l@#1\endcsname\relax
\typeout{** WARNING: IEEEtran.bst: No hyphenation pattern has been}%
\typeout{** loaded for the language `#1'. Using the pattern for}%
\typeout{** the default language instead.}%
\else
\language=\csname l@#1\endcsname
\fi
#2}}
\providecommand{\BIBdecl}{\relax}
\BIBdecl

\bibitem{Berrou1993}
C.~Berrou, A.~Glavieux, and P.~Thitimajshima, ``Near {S}hannon limit
  error-correcting coding and decoding: {T}urbo codes,'' in \emph{Proc. {IEEE}
  {ICC} 1993}, Geneva, Switzerland, May 1993, pp. 1064--1070.

\bibitem{Benedetto1998}
S.~Benedetto, D.~Divsalar, G.~Montorsi, and F.~Pollara, ``Serial concatenation
  of interleaved codes: performance analysis, design, and iterative decoding,''
  \emph{{IEEE} Trans. Inform. Theory}, vol.~44, no.~3, pp. 909--926, May 1998.

\bibitem{Bahl1974}
L.~Bahl, J.~Cocke, F.~Jelinek, and J.~Raviv, ``Optimal decoding of linear codes
  for minimizing symbol error rate,'' \emph{{IEEE} Trans. Inform. Theory},
  vol.~20, no.~3, pp. 284--287, Mar. 1974.

\bibitem{Gallager}
R.~G. Gallager, ``Low-density parity-check codes,'' \emph{IRE Trans. Inform.
  Theory}, vol. IT-8, pp. 21--28, Jan. 1962.

\bibitem{Richardson2001EfficientEncoding}
T.~Richardson and R.~Urbanke, ``Efficient encoding of low-density parity-check
  codes,'' \emph{{IEEE} Trans. Inform. Theory}, vol.~47, no.~2, pp. 638--656,
  Feb. 2001.

\bibitem{Pyndiah1998}
R.~Pyndiah, ``Near optimum decoding of product codes: block turbo codes,''
  \emph{{IEEE} Trans. Commun.}, vol.~46, no.~8, pp. 1003--1010, Aug. 1998.

\bibitem{Gazi2006}
O.~Gazi and A.~O. Yilmaz, ``Turbo product codes based on convolutional codes,''
  \emph{{ETRI} Journal}, vol.~28, no.~4, pp. 453--460, Aug. 2006.

\bibitem{RyanBook}
W.~E. Ryan and S.~Lin, \emph{Channel Codes - Classical and Modern}.\hskip 1em
  plus 0.5em minus 0.4em\relax Cambridge University, 2009.

\bibitem{Xu2005}
J.~Xu, L.~Chen, L.~Zeng, L.~Lan, and S.~Lin, ``Construction of low-density
  parity-check codes by superposition,'' \emph{{IEEE} Trans. Commun.}, vol.~53,
  no.~2, pp. 243--251, Feb. 2005.

\bibitem{Qi2004}
Z.~Qi and N.~C. Sum, ``{LDPC} product codes,'' in \emph{Proc. {ICCS} 2004},
  Kraków, Poland, Sep. 2004, pp. 481--483.

\bibitem{Esmaeili2006}
M.~Esmaeili, ``The minimal product parity check matrix and its application,''
  in \emph{Proc. {IEEE} {ICC} 2006}, Istambul, Turkey, Jun. 2006, pp.
  1113--1118.

\bibitem{Thomos2006}
N.~Thomos, N.~V. Boulgouris, and M.~G. Strintzis, ``Product code optimization
  for determinate state {LDPC} decoding in robust image transmission,''
  \emph{{IEEE} Trans. Image Processing}, vol.~15, no.~8, pp. 2113--2119, Aug.
  2006.

\bibitem{Hu2001PEG}
X.~Y. Hu and E.~Eleftheriou, ``Progressive edge-growth tanner graphs,'' in
  \emph{Proc. {IEEE} Global Telecommunications Conference ({GLOBECOM}'01)}, San
  Antonio, Texas, Nov. 2001, pp. 995--1001.

\bibitem{BaldiCL2009}
M.~Baldi, G.~Cancellieri, A.~Carassai, and F.~Chiaraluce, ``{LDPC} codes based
  on serially concatenated multiple parity-check codes,'' \emph{{IEEE} Commun.
  Lett.}, vol.~13, no.~2, pp. 142--144, Feb. 2009.

\bibitem{Chiaraluce2004}
F.~Chiaraluce and R.~Garello, ``Extended {H}amming product codes analytical
  performance evaluation for low error rate applications,'' \emph{{IEEE} Trans.
  Wireless Commun.}, vol.~3, no.~6, pp. 2353--2361, Nov. 2004.

\bibitem{Baldi2009}
M.~Baldi, G.~Cancellieri, and F.~Chiaraluce, ``A class of low-density
  parity-check product codes,'' in \emph{Proc. {SPACOMM} 2009}, Colmar, France,
  Jul. 2009, pp. 107--112.

\bibitem{Roth2006}
R.~M. Roth, \emph{Introduction to Coding Theory.}\hskip 1em plus 0.5em minus
  0.4em\relax Cambridge University Press, 2006.

\bibitem{Hagenauer1996}
J.~Hagenauer, E.~Offer, and L.~Papke, ``Iterative decoding of binary block and
  convolutional codes,'' \emph{{IEEE} Trans. Inform. Theory}, vol.~42, no.~2,
  pp. 429--445, Mar. 1996.

\bibitem{Perez1996}
L.~C. Perez, J.~Seghers, and D.~J. Costello, ``A distance spectrum
  interpretation of turbo codes,'' \emph{{IEEE} Trans. Inform. Theory},
  vol.~42, no.~6, pp. 1698--1709, Nov. 1996.

\bibitem{802.16e}
802.16e 2005, \emph{{IEEE} {S}tandard for {L}ocal and {M}etropolitan {A}rea
  {N}etworks - {P}art 16: {A}ir {I}nterface for {F}ixed and {M}obile
  {B}roadband {W}ireless {A}ccess {S}ystems}, {IEEE} Std., Dec. 2005.

\end{thebibliography}

\end{document}